\documentclass[aps,prl,showpacs,twocolumn,groupedaddress]{revtex4-1}
\usepackage{amsmath}
\usepackage{amsfonts}
\usepackage{amssymb}
\usepackage{graphicx}
\usepackage{verbatim}
\usepackage{xcolor}
\usepackage[normalem]{ulem}
\newcommand{\ts}{\textsuperscript}

\begin{document}

\title{High harmonic generation tomography of impurities in solids: conceptual analysis}

\author{S. Almalki\ts{1}}
\email[]{salma012@uottawa.ca}
\author{A. Parks\ts{1}}
\author{G. Bart\ts{1}}
\author{P. B. Corkum\ts{1,2}}
\author{T. Brabec\ts{1}}
\author{C. R. McDonald\ts{1}}
\email[]{cmcdo059@uottawa.ca}

\affiliation{\ts{1}Department of Physics, University of Ottawa, Ottawa, ON K1N 6N5, Canada}
\affiliation{\ts{2}National Research Council of Canada, Ottawa, Ontario, Canada K1A 0R6}

\date{\today}

\begin{abstract}
A three step model for high harmonic generation from impurities in solids is developed. The process is found to be similar to high harmonic generation in atomic and molecular gases with the main difference coming from the non-parabolic nature of the bands. This opens a new avenue for strong field atomic and molecular physics in the condensed matter phase. As a first application, our conceptual study demonstrates the feasibility of tomographic measurement of impurity orbitals. 

\end{abstract}

\maketitle

\noindent
Strong field and attosecond science in atomic and molecular physics has made great strides over the past 20 years \cite{Corkum1993, Corkum2007}. Strong laser-atom interaction takes place in a three step process; first the weakest 
bound electron is ionized, followed by laser driven evolution in the continuum, and finally it recollides/rescatters with its parent ion \cite{Corkum1993, Lewenstein1994}. It has been found that both tunneling and recollision processes 
contain a great deal of information about the parent system's structure and dynamics. 

Angular resolved tunnel ionization spectroscopy \cite{Tong2002,Akagi2009} reveals the orbital angular structure of the highest occupied molecular orbital. 

When the electron recollides with its parent ion, recombination and rescattering take place. Recombination results in high harmonic generation (HHG) --- the emission of coherent XUV radiation. HHG has been used to time resolve chemical 
reactions and to tomographically measure the wavefunction of simple molecules \cite{Itatani2004, Baker2006, Worner2010, Shafir2009, Patchkovskii2006, Vozzi2011}. 

Rescattering \cite{Lin2010} results in nonsequential double ionization, above threshold ionization, and laser induced electron diffraction \cite{Fittinghoff1992, Gaarde2000, Walker1996, Wolter2016}; these processes have structural 
information encoded and are also promising candidates for time resolved imaging of molecular reactions. 

Recent experiments with mid-infrared \cite{Ghimire2011, Luu2015, Vampa2015_1, Ndabashimiye2016} and THz pump sources \cite{Zaks2012, Schubert2014, Hohenleutner2015} have demonstrated HHG in solids. Theory has identified two mechanisms 
\cite{Golde2008, Vampa2014}; (i) intraband HHG due to the non-parabolic nature of bands \cite{Luu2015} was found to be dominant in dielectrics; (ii) interband HHG dominates in semiconductors and is created in a three step process similar 
to atomic and molecular HHG \cite{Vampa2015_2}. This similarity has established a connection between attosecond physics in atoms/molecules and in the condensed matter phase.

Our analysis further deepens the links between strong field physics in the gas and condensed matter phases. Recently, HHG from impurities in solids has been demonstrated \cite{Sivis2017, Huang2017}. We develop quantum equations of motion 
and a three step model for this process. First, a free electron/hole is created in the conduction/valence band by tunnel ionization of a donor/acceptor impurity. Second, the electron/hole is accelerated by the laser field. In a third step 
a harmonic photon is emitted upon recollision and recombination with the parent impurity. 

Besides differences in the continuum evolution due to the non-parabolic nature of bands, the process is found to be identical to HHG in gases. As a consequence, many of the above processes can be adapted from the gas to the condensed 
matter phase. This opens a new research direction for atomic and molecular strong field processes. 

As a first application, we study the potential of applying molecular HHG tomography \cite{Shafir2009, Patchkovskii2006, Vozzi2011} to impurities. Tomographic reconstruction of the impurity ground state is demonstrated in a 1D model system. 
The impurity dipole moment is found to be the dominant factor in determining the magnitude of the harmonic signal as a function of harmonic order; ionization and propagation which have to be factored out in molecular tomography play a 
lesser role here. This indicates a substantial facilitation due to the potential for direct reconstruction of the impurity ground state from the harmonic spectrum. 

Our results create a link between strong field physics and solotronics --- solitary impurity electronics; for a review see Ref. \onlinecite{Koenraad2011}. Solitary impurities are important building blocks for quantum technology --- 
as qubits for quantum computing and as single photon and non-classical photon sources for quantum sensing and communication. Further, with increasing miniaturization the device characteristics of MOSFET transistors is increasingly 
stronger influenced by scattering off single impurities. All of the above applications require detailed knowledge about the wavefunction of impurity and environment. Currently, the most powerful method to image the wavefunction of 
single impurities is scanning tunneling microscopy close to suitably cleaved surfaces. Our results reveal that strong field methods can offer complementary capacities. Among other things they provide an all-optical way to measure 
dipole moment and wavefunctions of impurity ensembles independent of surfaces; single impurity imaging will be challenging due to the low quantum yield of HHG. Beyond that they open the path to spatio-temporal imaging of wavefunction dynamics in impurities, impurity molecules and arrays \cite{Koenraad2011} via optical pump probe experiments. 

Our analysis is based on the following model: an impurity with potential $U({\bf x})$ is imbedded in a solid and is coupled to a laser field $\mathbf{F}(t)$ via the dipole coupling term $\mathbf{x}\cdot\mathbf{F}(t)$. The resulting 
Hamiltonian is given by 
\begin{align}
i \partial_t \Psi(\mathbf{x},t) = \left( H_i - \mathbf{x}\cdot\mathbf{F}(t) \right) \Psi(\mathbf{x},t) \text{,}
\label{hamiltonian} 
\end{align}
where $H_i = H_0 + U({\bf x})$ and $H_0$ refers to the Hamiltonian of the solid without impurity. Atomic units, $e = \hbar = m_e = 1$, are used throughout unless otherwise indicated. 

Shallow donor (acceptor) impurities split into an electron (hole) and a positively (negatively) charged residual ion; the electron (hole) moves in the lowest conduction (highest valence) band and has bound states in the field of the residual ion with energies closely below (above) the bottom (top) of the conduction (valence) band. As a result, we focus our treatment to a single band with eigenfunctions $\Phi_{{\bf k}}$ which fulfil $H_0 \Phi_{{\bf k}} = E({\bf k}) \Phi_{{\bf k}}$ with $E({\bf k})$ the band eigenenergies; further, $\Phi_{{\bf k}}({\bf x}) = u_{{\bf k}}({\bf x}) \exp(i{\bf k}\!\cdot\!{\bf x})$ with $u_{{\bf k}}$ the Bloch function that is periodic with the lattice. The crystal momentum $\mathbf{k}$ extends over the first Brillouin zone (BZ) defined by the unit cell lattice vectors $\mathbf{a}_l$ and reciprocal lattice vectors $\mathbf{b}_l$ with $|\mathbf{b}_l| = \pi / |\mathbf{a}_l|$ and $l=1,2,3$. 

The ground state of shallow impurities is derived by following standard theory \cite{Luttinger1955, Kohn1955_1, Pantelides1978, Yu2001}; a derivation is given in the Supplementary Material \cite{som}.   The impurity ground state is given by, 
\begin{align}
\phi_0({\bf x}) = \Phi_{\mathbf{k}=0}({\bf x}) B_0({\bf x})
\label{effmground} 
\end{align}
where $B_0$ is the ground state envelope with energy $\varepsilon_0$, as determined by 
\begin{align}
(\varepsilon_0 - E_g) B_0({\bf x}) = \sum_{j,l} \frac{\beta_{jl}}{2} \mathbf{\nabla}_j \mathbf{\nabla}_l B_0({\bf x}) + U({\bf x}) B_0({\bf x}) \text{.}
\label{effmeq} 
\end{align}
Both envelope $B_0$ and $U$ are assumed to vary slowly over one unit cell. Here, $j,l = x,y,z$ and $\beta_{jl}$ is the inverse mass tensor that arises from the quadratic expansion of the band energy $E({\bf k})$ about the $\Gamma$-point 
($\mathbf{k}=0$), where the band energy $E(\mathbf{k}=0) = E_g$ is minimum; for the sake of simplicity we confine our analysis to direct bandgap materials; generalization to indirect bands can be done 
by following the treatment in Ref. \onlinecite{Kohn1955_1}. 

The time-dependent Schr\"odinger equation (\ref{hamiltonian}) is solved by splitting the wavefunction into a bound state part and a band contribution, 
\begin{align}
\Psi({\bf x},t)  = \phi_0({\bf x}) + \int_{\rm BZ} \!\!\! a({\bf k}, t) \Phi_{\mathbf{k}}({\bf x}) \, d^{3}{\bf k} \text{,}
\label{ansatz1} 
\end{align}
where the integral runs over the first BZ. We assume, in the spirit of the strong field approximation \cite{Lewenstein1994}, that field induced ionization is weak enough so that the ground state population remains unaffected. 
This amounts to neglecting the dynamic Stark shift of the impurity ground state. 

Next Ansatz (\ref{ansatz1}) is inserted into Eq. (\ref{hamiltonian}) and the resulting equation is multiplied with $\langle \Phi_{\mathbf{k}'} |$. This yields an equation of motion for the probability amplitude of the 
conduction band states, 
\begin{align}
i \frac{d}{dt}a({\bf k},t) & = \left( E({\bf k}) + i {\bf F}(t)\cdot\mathbf{\nabla}_{\bf k} \right) a({\bf k},t) + i \bf {d}_0({\bf k})\cdot{\bf F}(t) \nonumber \\
~ & + \int U({\bf k} - {\bf k}') a({\bf k}', t) d^3{\bf k}'  \text{.} 
\label{eqmot} 
\end{align}
Again a soft impurity potential is assumed for which large momentum scattering $|\mathbf{k}| > |\mathbf{b}_l|$ is negligible \cite{som}. Here, the transition dipole moment between impurity ground state and conduction band is  ${\bf d}_0({\bf k}) = \langle \Phi_{\mathbf{k}} | {\bf x} | \phi_0 \rangle$. The bra-ket notation implies integration over the whole crystal volume $V$. As long as impurity ground state and conduction band wavefunction vary slowly compared to the Bloch functions, the dipole moment becomes,
\begin{align}
{\bf d}_0({\bf k}) \approx (2\pi)^{-3}\!\! \int_{V} \mathbf{x} B_0(\mathbf{x})e^{-i\mathbf{k}\cdot\mathbf{x}} d^{3}{\bf x} \text{,}
\label{dipole} 
\end{align}
which is formally identical with the dipole moment of atomic gases. Finally, agreement with atomic strong field physics becomes complete, when the quadratic mass approximation is applied to Eq. (\ref{eqmot}) 
\cite{Keldysh1965, Lewenstein1994}. 

In the strong field limit, the Coulomb potential in Eq. (\ref{eqmot}) is neglected. For impurities additional justification comes from the fact that photoionization cross sections are well described by replacing the Coulomb with 
delta-function potentials \cite{Pantelides1978}. Integration of the resulting Eq. (\ref{eqmot}) and inserting the result into the the second term of Eq. (\ref{ansatz1}) yields the time dependent evolution of the electron wavefunction 
in the conduction band as
\begin{align}
a({\bf k},t) \! = \! \int_{-\infty}^{t} \!\!\!\!\! dt' {\bf d}_0(\boldsymbol{\kappa}_{t'}) \!\cdot\! {\bf F}(t') e^{ \int_{-\infty}^{t'} \! i \big( \varepsilon_0 - 
E(\boldsymbol{\kappa}_{t''})  + \frac{i}{T_2}\big) dt'' }
\label{tdwf} 
\end{align}
where $\boldsymbol{\kappa}_{t'} = {\bf k} - {\bf A}(t) + {\bf A}(t')$ with vector potential determined by ${\bf F} = -d{\bf A}/dt$; further, a phenomenological dephasing time $T_2$ has been added. 

The high harmonic electric fields are determined by the polarization buildup between band and impurity ground state resulting in a current 
\begin{align}
\tilde{j}_{i}(\omega) & = i\omega \int_{\rm BZ} \!\! d^3\!{\bf k} \, {\bf d}_0^*({\bf k}) \int_{-\infty}^{\infty} \!\! dt e^{- i \omega t} \!\!\int_{-\infty}^{t} \!\!\! dt' 
{\bf d}_0(\boldsymbol{\kappa}_{t'})\!\cdot\! {\bf F}(t')
\nonumber \\
~ & \times  e^{\big( i S({\bf k},t',t) - \frac{1}{T_2}(t-t')\big) dt'' }  + \text{c.c.}
\label{recdip} 
\end{align}
where $S({\bf k},t',t) = \int_{t'}^{t}\!\big(\varepsilon_0 - E(\boldsymbol{\kappa}_{t''})\big)dt''$.

The three integrals in Eq. (\ref{recdip}) can be solved analytically by the saddle point method \cite{Lewenstein1994}. The saddle point equations are determined by, 
\begin{subequations}
\label{spall}
\begin{align}
\nabla_{\bf k} S & =  \int_{t'}^t {\bf v}(\boldsymbol{\kappa}_{t''}) dt'' = {\bf x}(t) - {\bf x}(t') = 0 & \label{sp1} \\
{dS \over dt'}  & = E({\bf k} - {\bf A}(t) + {\bf A}(t')) - \varepsilon_0 = 0 & \label{sp2} \\
{dS \over dt} & = E({\bf k}) - \varepsilon_0 = \omega \text{.} \label{sp3} &
\end{align}
\end{subequations}
In Eq. (\ref{sp1}), the band velocity is given by ${\bf v}({\bf k}) = \boldsymbol{\nabla}_{\bf k} E$. This equation states that HHG can take place only when the electron, born at time $t'$ into the band, returns to the parent impurity 
at $t$. Equation (\ref{sp2}) states that electrons are born with zero momentum at time $t'$, ${\bf k} = {\bf A}(t=t') - {\bf A}(t') = 0$. At the time of recombination $t$ the electron crystal momentum is given by ${\bf k}(t',t) = {\bf A}(t) - {\bf A}(t')$. The finite impurity gap energy results in a complex birth time, which is responsible for tunnel ionization. Finally, Eq. (\ref{sp3}) represents conservation of energy --- the electron recombines to the ground state and emits a photon $\omega$ with energy equal $E({\bf k(t',t)}) - \varepsilon_0$. Again, at moderate laser intensities, for which the effective mass approximation applies, the saddle point equations for atom and impurity become identical. 

After saddle point integration we obtain for the harmonic intensity 
\begin{align}
\left| \tilde{j}_{i}(\omega) \right|^2 \!  & = \! \left| \sum_{t'} \! \sqrt{\mathrm{w}(t')} \mathbf{d}_{0}^*\!(\mathbf{k}) \alpha(t',t) 
e^{ \int_{t'}^{t} \!\big(i \mathcal{S} - \frac{1}{T_2}\big) dt'' } \right| ^2 \text{,}
\label{recdips} 
\end{align}
where $\mathcal{S} = \varepsilon_0 + \omega - E\big( \mathbf{k}(t'',t) \big)$, $\mathrm{w}(t')$ is the ionization rate and $t'(t(\omega))$ and $t(\omega)$ are birth and recombination times resulting in the generation of a harmonic with frequency $\omega$. For an isotropic lattice the ionization rate is determined by the ADK tunnel ionization rate of atoms \cite{Tong2002} with the electron mass replaced by the effective mass. Further, the dipole moment represents the recombination amplitude; the remaining term $\alpha$ in the pre-exponential is the propagation amplitude accounting for quantum diffusion and dephasing; this depends on the band specifics. For isotropic materials in the effective mass $m$ approximation $\alpha \propto m\exp(-(t-t')/T_2)(t-t')^{-3/2}$. The main difference between HHG from impurities and atoms arises from the finite, non-parabolic, anisotropic nature of bands. 

In the remaining part we use the above formalism to investigate HHG and the tomographic reconstruction of the impurity ground state wavefunction from harmonic spectra. We use a 1D model system for a direct band gap semiconductor.  The periodic lattice potential is composed of lattice cells of width $a = 9.45\,{\rm a.u.}$ and well depth $v_0 = 0.55\,{\rm a.u.}$  The lattice cells are separated by a mollifier function \cite{McDonald2017}. The impurity Coulomb potential is $U(x) = -1/({\epsilon} \sqrt{x^2+s^2} )$ where $s = 25\,{\rm a.u.}$ is the softening parameter and $\epsilon = 5\,{\rm a.u.}$ is the dielectric constant. The resulting Hamiltonian is then diagonalized, both with and without $U(x)$, using periodic boundary conditions.  In the absence of $U(x)$ we obtain the Bloch states for the unperturbed system and with $U(x)$ present, we obtain the impurity ground state. For this system the impurity ground state is such that $E_g - \varepsilon_0 = 3.9 \times 10^{-3}\,{\rm a.u.}$ ($106.4\,{\rm meV}$). The system is irradiated by a laser field  with vector potential $A(t) = - (F_0/\omega_0) f(t) \sin(\omega_0 t)$ with peak field strength $F_0= 1 \times 10^{-4} {\rm a.u.}$ ($I_0 = 3.5\times 10^{8}\,{\rm W/cm}^2$) and center frequency $\omega_0 = 9.1\times 10^{-4}\,{\rm a.u.}$ ($\lambda_0 = 50\,\mu {\rm m}$). The  envelope $f(t)$ is Gaussian with a FWHM of 12 laser cycles. The time dynamics of the system are determined from Eq. (\ref{tdwf}) with $T_2 = 50\,{\rm fs}$. 

\begin{figure}
\includegraphics[scale=1.0]{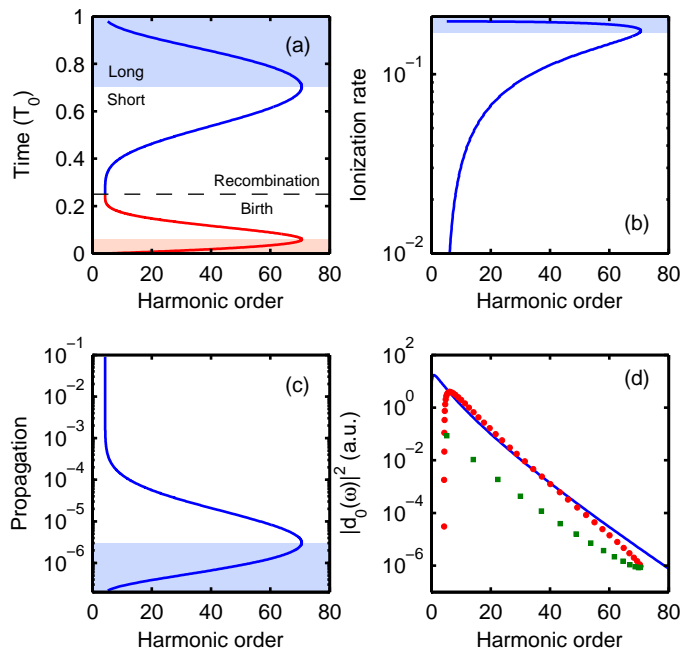}
\caption{\label{fig1} (a) Birth time $t'$ (red) and return times $t$ (blue) from the semiclassical trajectories versus harmonic order. (b) Ionization rate ${\rm w} \propto \exp \big(-\frac{2}{3}\sqrt{m} (2(E_g - \varepsilon_0))^{3/2}/F(t')\big)$ versus harmonic order. (c) Propagation effects $\alpha^2 \propto \exp\big(-2(t - t')/T_2\big) / (t - t')$ versus harmonic order. (d) Magnitude squared of the dipole moment as a function of harmonic order (blue); the product of the three pre-exponential terms in Eq. (\ref{recdips}) represented by blue lines in (b)-(d) is plotted for the short (red dots) and long (green squares) trajectory branches; the magnitude is adapted to match the dipole moment.  In (a) - (c) the shaded regions indicate the contributions from long trajectories.}
\end{figure}

\begin{figure}
\includegraphics[scale=1.0]{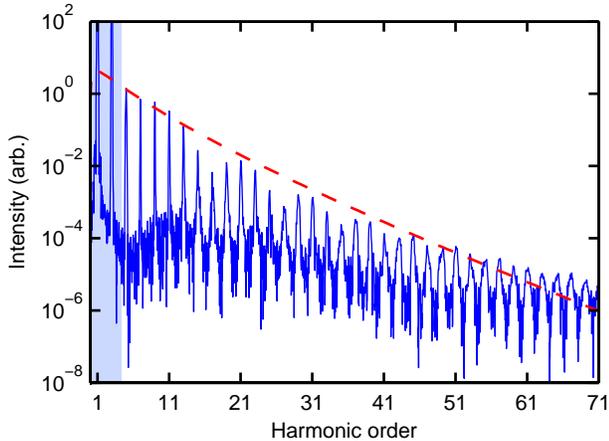}
\caption{\label{fig2} Scaling comparison of the harmonic spectrum (blue) to the dipole (red, dashed); the dipole has been shifted down in order to compare with the spectrum. The harmonics in the shaded region are those whose energy 
is below $E_g - \varepsilon_0$.}
\end{figure}

Figure \ref{fig1}(a) shows the generated harmonics versus birth (red) and recombination (blue) times from the semiclassical trajectories obtained from numerical solution of Eqs. (\ref{spall}). There are two sets of solutions per optical cycle, a short and a long trajectory. The long trajectory contributions are indicated by the shaded regions in Fig. \ref{fig1}. Figures \ref{fig1}(b)-(d) examine the behavior of each of the pre-exponential terms in Eq. (\ref{recdips}). Figures \ref{fig1}(b) and \ref{fig1}(c) present the ionization rate and propagation term, respectively. For ionization we have used the dominant atomic tunneling exponent \cite{Keldysh1965, Lewenstein1994}. Figure \ref{fig1}(d) shows $|d_0(\omega)|^2$ obtained from the diagonalization of the Hamiltonian (blue line), where $k$ has been replaced with $\omega$ by virtue of relation (\ref{sp3}). We find that $|d_0(\omega)|^2$ decreases by about six orders of magnitude with increasing harmonic order. The rapid drop comes from the fact that the ground state extends over many unit cells and therewith populates only a small fraction of the BZ. In Fig. \ref{fig1}(d) we also plot the product of all three terms, where long and short trajectories are indicated by red dots and green squares, respectively. The short trajectories are dominant and a comparison with $|d_0(\omega)|^2$ shows that the dipole moment determines the form of the harmonic spectrum over most of its range; this is confirmed in Fig. \ref{fig2}. 

In Fig. \ref{fig2} the harmonic intensity $|\tilde{j}|^2$ (blue) is plotted, including both the impurity and intraband contributions.  We note that the harmonics above the impurity ionization potential are dominated by the impurity term (see Supplementary Material \cite{som}). The strength of the above impurity gap harmonics drops rapidly until around the cut-off near the $71^{st}$ harmonic. This behavior is consistent with the decrease of $|d_0(\omega)|^2$ (dashed) indicating that, of the three pre-exponential terms in Eq. (\ref{recdips}), the dipole has the strongest influence on the shape of the harmonic spectrum. Consequently, using relation (\ref{sp3}) to connect harmonic order and $k$, we can reconstruct $d_0(k)$ from the magnitude of the harmonic spectrum. This is feasible, as the atom-like dipole moment is purely real/imaginary. For a complex dipole moment the phase of the harmonics must be considered, as in reference \onlinecite{Boutu2008}. 

\begin{figure}
\includegraphics[scale=1.0]{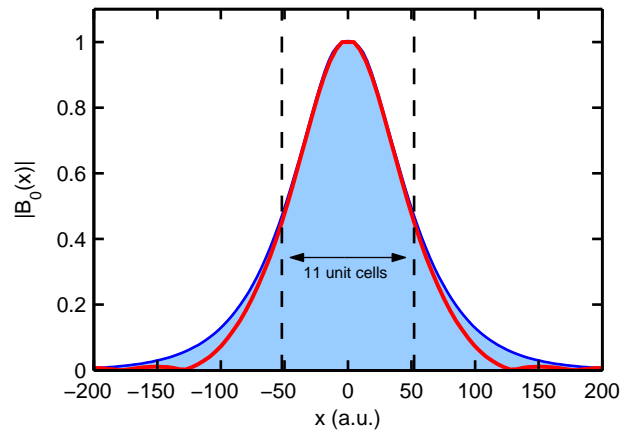}
\caption{\label{fig3} Comparison between the impurity ground state (blue, shaded) and the reconstructed ground state (red). The region between the vertical dashed lines represents 11 unit cells.}
\end{figure}

To reconstruct the impurity ground state we take the inverse Fourier transform of $d_0(k)$ and divide it by $x$ to obtain $B_0(x)$. In a 3D experiment one would rotate the crystal and reconstruct the total wavefunction from 1D snapshots. Figure \ref{fig3} shows the results of the tomographic reconstruction. The reconstructed wavefunction (red) matches the impurity ground state well throughout the central region but deviates from the true wavefunction at the tails. This agrees with the fact that the difference between harmonic intensity and dipole scaling in Fig. \ref{fig2} is biggest for small crystal momenta corresponding to slow wavefunction variations in real space. Further the small oscillations in the harmonic spectrum in Fig. \ref{fig2} do not appear to cause a substantial error in the reconstruction; they result from interference between harmonics generated in positive and negative half-cycles as a consequence of the phase term in Eq. (\ref{recdips}). 

The approach developed here presents a theoretical underpinning for exploring strong field physics in impurities and for adapting technology developed for atomic and molecular gases to solid state impurities. In contrast to gases, absorption will limit the material depth from which photons and electrons can be detected; as a result, propagation effects are expected to be less significant. Whether experimental tomographic reconstruction is as straightforward as found here remains to be seen. How dominant the dipole moment is in determining harmonic spectra will depend on various factors, such as dephasing time, material dimension and parameters. In the Supplement \cite{som} a 3D impurity is discussed which underlies stronger quantum spreading. As a result, quantum diffusion would need to be factored out of the spectrum for a more accurate reconstruction; this is straightforward using Eq. (\ref{recdips}). In some systems more sophisticated reconstruction techniques will be required. For example, the ionization rate could be measured by transient absorption spectroscopy \cite{Schultze2014} and factored out of the harmonic spectrum. Furthermore, the dipole moment extracted from Eq. (\ref{recdips}) or from numerical analysis can be improved on by using optimization techniques, similar to the one used recently for all optical band gap measurements \cite{Vampa2015}. 

Finally, it needs to be discussed that our theoretical approach has been developed for shallow impurities. Deep impurities are more complex, as their wavefunction changes substantially over a unit cell. This results in a strong mixing between impurity and surrounding lattice wavefunction components. The resulting many-body effects, such as coupling to quasi-particles, need to be addressed with more sophisticated theoretical approaches \cite{Koenraad2011}. They will dominantly enter in the dipole moment and therewith in recombination; ionization will be influenced to a lesser extent, as the dipole moment enters in the pre-exponent. Propagation will only be altered close to the impurity, where the impurity potential yields higher-order corrections to the strong field approximation. As a result, our simple approach will present a reasonable starting point to develop strong field impurity physics in this more complex limit.



\begin{thebibliography}{99}

\bibitem{Corkum1993} P. B. Corkum, {\it Plasma perspective on strong field multiphoton ionization}, \prl {\bf 71}, 1994 (1993).
\bibitem{Corkum2007} P. B. Corkum and F. Krausz, {\it Attosecond science}, Nat. Phys. {\bf 3}, 381 (2007).
\bibitem{Lewenstein1994} M. Lewenstein, P. Balcou, M. Y. Ivanov, A. L'Huillier, P. B. Corkum, {\it Theory of high-harmonic generation by low-frequency laser fields}, 
\pra \textbf{49}, 2117 (1994).
\bibitem{Tong2002} X. M. Tong, Z. X. Zhao and C. D. Lin, {\it Theory of molecular tunneling ionization}, \pra \textbf{66}, 033402 (2002).
\bibitem{Akagi2009} H. Akagi, T. Otobe, A. Staudte, A. Shiner, F. Turner, R. D\"orner, D. M. Villeneuve, and P. B. Corkum, {\it Laser tunnel ionization from muiltiple
orbitals in HCl}, Science \textbf{325}, 1364 (2009). 
\bibitem{Itatani2004} J. Itatani, J. Levesque, D. Zeidler, H. Niikura, H. P{\'e}pin, J. C. Kieffer, P. B. Corkum and D. M. Villeneuve, \textit{Tomographic imaging of
molecular orbitals}, Nature \textbf{432}, 867 (2004).
\bibitem{Baker2006} S. Baker, J. S. Robinson, C. A. Haworth, H. Teng, R. A. Smith, C. C. Chirila, M. Lein, J. W. G. Tisch and J. P. Marangos, {\it Probing Proton
Dynamics in Molecules on an Attosecond Time Scale}, Science {\bf 312}, 424 (2006). 
\bibitem{Worner2010} H. J. W\"orner, J. B. Bertrand, D. V. Kartashov, P. B. Corkum and D. M. Villeneuve, {\it Following a chemical reaction using high-harmonic
interferometry}, Nature {\bf 466}, 604 (2010). 
\bibitem{Shafir2009} D. Shafir, Y. Mairesse, D. Villeneuve, P. B. Corkum and N. Dudovich, {\it Atomic wavefunctions probed through strong-field light–matter 
interaction}, Nat. Phys. {\bf 5}, 412 (2009). 
\bibitem{Patchkovskii2006} S. Patchkovskii, Z. Zhao, T. Brabec and D. M. Villeneuve, \textit{High Harmonic Generation and Molecular Orbital Tomography in 
Multielectron Systems: Beyond the Single Active Electron Approximation}, \prl \textbf{97}, 123003 (2006).
\bibitem{Vozzi2011} C. Vozzi, M. Negro, F. Calegari, G. Sansone, M. Nisoli, S. De Silvestri and S. Stagira, \textit{Generalized molecular orbital tomography}, 
Nature Phys. \textbf{7}, 822 (2011).
\bibitem{Lin2010} C. D. Lin, Anh-Thu Le, Z. Chen, T. Morishita, and R. Lucchese, {\it Strong-field rescattering physics—self-imaging of a molecule by its own 
electrons}, J. Phys. B \textbf{43}, 122001 (2010). 
\bibitem{Wolter2016} B. Wolter, M. G. Pullen, A.-T. Le, M. Baudisch, K. Doblhoff-Dier, A. Senftleben, M. Hemmer, C. D. Schröter, J. Ullrich, T. Pfeifer, R. Moshammer, 
S. Gräfe, O. Vendrell, C. D. Lin and J. Biegert, {\it Ultrafast electron diffraction imaging of bond breaking in di-ionized acetelyne}, Chem. Phys. \textbf{354}, 
308 (2016). 
\bibitem{Fittinghoff1992} D. N. Fittinghoff, P. R. Bolton, B. Chang, and K. C. Kulander, {\it Observation of nonsequential double ionization of helium with optical
tunneling}, \prl \textbf{69}, 2642 (1992). 
\bibitem{Gaarde2000} M. B. Gaarde, K. J. Schafer, K. C. Kulander, B. Sheehy, D. Kim, L. F. DiMauro, Phys. Rev. Lett. \textbf{84} 2822 (2000).  
\bibitem{Walker1996} B. Walker, B. Sheehy, K. C. Kulander, and L. F. DiMauro, {\it Elastic Rescattering in the Strong Field Tunneling Limit}, \prl \textbf{77}, 5031 
(1996). 
\bibitem{Ghimire2011} S. Ghimire \textit{et al.}, \textit{Observation of high-order harmonic generation in a bulk crystal}, {Nat. Phys.} \textbf{7}, 138 (2011). 
\bibitem{Luu2015} T. T. Luu, M. Garg, S. Yu. Kruchinin, A. Moulet, M. Th. Hassan and E. Goulielmakis, \textit{Extreme ultraviolet high-harmonic spectroscopy of 
solids}, {Nature} \textbf{521}, 498 (2015).  
\bibitem{Vampa2015_1} G. Vampa, T. J. Hammond, N. Thir\'{e}, B. E. Schmidt, F. L\'{e}gar\'{e}, C. R. McDonald, T. Brabec, and P. B. Corkum \textit{Linking high 
harmonic generation from gases and solids}, {Nature} \textbf{522}, 462 (2015). 
\bibitem{Ndabashimiye2016} G. Ndabashimiye, S. Ghimire, M. Wu, D. A. Browne, K. J. Schafer, M. B. Gaarde and D. A. Reis {\it Solid-state harmonics beyond the 
atomic limit}, Nature {\bf 534}, 520 (2016). 
\bibitem{Zaks2012} B. Zaks, R. B. Liu and M. S. Sherwin, \textit{Experimental observation of electron–hole recollisions}, {Nature} \textbf{483}, 580 (2012). 
\bibitem{Schubert2014} O. Schubert et al. {\it Sub-cycle control of terahertz high-harmonic generation by dynamical Bloch oscillations}, Nat. Photon. {\bf 8}, 
119 (2014). 
\bibitem{Hohenleutner2015} M. Hohenleutner \textit{et al.}, \textit{Real time observation of interfering crystal electrons in high-harmonic generation}, {Nature} 
\textbf{523}, 572 (2015). 
\bibitem{Golde2008} D. Golde, T. Meier and S. W. Koch, {\it High harmonics generated in semiconductor nanostructures by the coupled dynamics of optical inter- and
intraband excitations}, \prb {\bf 77}, 075330 (2008).
\bibitem{Vampa2014} G. Vampa, C. R. McDonald, G. Orlando, D. D. Klug, P. B. Corkum, and T. Brabec, \textit{Theoretical Analysis of High-Harmonic Generation in Solids},
\prl {\bf 113}, 073901 (2014). 
\bibitem{Vampa2015_2} G. Vampa, C. R. McDonald, G. Orlando, P. B. Corkum and T. Brabec, {\it Semiclassical analysis of high harmonic generation in bulk crystals}, 
\prb {\bf 91}, 064302 (2015). 
\bibitem{Sivis2017} M. Sivis, M. Taucer, G. Vampa, K. Johnston, A. Staudte, A. Yu Naumov, D. M. Villeneuve, C. Ropers, and P. B. Corkum, \textit{Tailored semiconductors for high-harmonic optoelectronics}, Science \textbf{357}, 303 (2017).
\bibitem{Huang2017} T. Huang, X. Zhu, L. Li, X. Liu, P. Lan, and P. Lu, {\it High-order-harmonic generation of a doped semiconductor}, \pra {\bf 96}, 043425 (2017). 

\bibitem{Koenraad2011} P. M. Koenraad and M. E. Fratt\'e, \textit{Single dopands in semiconductors}, Nat. Mat. \textbf{10}, 91 (2011). 


\bibitem{Luttinger1955} J. M. Luttinger and W. Kohn, \textit{Motion of Electrons and Holes in Perturbed Periodic Fields}, Phys. Rev. {\bf 97}, 869 (1955). 
\bibitem{Kohn1955_1} W. Kohn and J. M. Luttinger, \textit{Hyperfine Splitting of Donor States in Silicon}, Phys. Rev. {\bf 97}, 883 (1955). 
\bibitem{Pantelides1978} S. T. Pantelides, {\it The electronic structure of impurities and other point defects in semiconductors}, \rmp \textbf{50}, 797 (1978). 
\bibitem{Yu2001} P. Y. Yu and M. Cardona, Fundamentals of Semicondcutors, Springer-Verlag Berlin Heidelberg (2001). 
\bibitem{som} Supplementary Material http://
\bibitem{Keldysh1965} L. Keldysh, \textit{Ionization in the field of a strong electromagnetic wave}, Sov. Phys. JETP \textbf{20}, 1307 (1965). 
\bibitem{McDonald2017} C. R. McDonald, K. S. Amin, S. Almalki, and T. Brabec, \textit{Enhancing High Harmonic Output in Solids through Quantum Confinement}, 
\prl {\bf 119}, 183902 (2017).
\bibitem{Boutu2008} W. Boutu, S. Haessler, H. Merdji, P. Breger, G. Waters, M. Stankiewicz, L. J. Frasinski, R. Taieb, J. Caillat, A. Maquet, P. Monchicourt, 
B. Carre, and P. Salieres, \textit{Coherent control of attosecond emission from aligned molecules}, Nat. Phys. \textbf{4}, 545 (2008). 
\bibitem{Schultze2014} M. Schultze, K. Ramasesha, C.D. Pemmaraju, S.A. Sato, D. Whitmore, A. Gandman, J. S. Prell, L. J. Borja, D. Prendergast, K. Yabana, 
D. M. Neumark, S. R. Leone, \textit{Attosecond bandgap dynamics in silicon}, Science \textbf{346}, 1348 (2014). 
\bibitem{Vampa2015} G. Vampa, T. J. Hammond, N. Thir\'e, B. E. Schmidt, F. Legar\'e, C. R. McDonald, T. Brabec, D. D. Klug, and P. B. Corkum, \textit{All-optical reconstruction of crystal bandstructure}, \prl \textbf{115}, 193603 (2015). 

\end{thebibliography}
\end{document}